\documentstyle[12pt]{article}
\begin{document}
\baselineskip=21pt
\begin{titlepage}
\rightline{Alberta Thy-8-94}
\rightline{MARCH 1994}
\vskip .1in
\begin{center}
{\large{\bf  FIELD REDEFINITIONS IN STRING THEORY AS

A SOLUTION GENERATING TECHNIQUE}}
\end{center}

\vskip .1in
\begin{center}

Nemanja Kaloper
\vskip.2in
Theoretical Physics Institute
\vskip.05in
Department of Physics, University of Alberta
\vskip.05in
Edmonton, Alberta, Canada T6G 2J1
\vskip.05in
{\it email: kaloper@fermi.phys.ualberta.ca}
\vskip.2in
\vskip.5cm
\end{center}
\centerline{ {\bf Abstract} }
\baselineskip=18pt

The purpose of this work is to show that there exists an additional invariance
of the $\beta$-function equations of
string theory on $d+1$-dimensional targets with $d$ toroidal isometries.
It corresponds to a shift of the
dilaton field and a scaling of the lapse function,
and is reminiscent of string field redefinitions. While it
preserves the form of the  $\beta$-function equations, it changes
the effective action and
the solutions. Thus it can be used as a solution generating technique.
It is particularly interesting to note
that there are field redefinitions which map solutions with non-zero
string cosmological constant to those with zero cosmological constant.
Several simple examples involving two- and three-dimensional black holes
and black strings are provided to illustrate the role of such field
redefinitions.

\vskip1.5cm
\centerline{\it Submitted to Phys. Lett. B}
\end{titlepage}

\baselineskip=21pt

\newpage
{\newcommand{\la}{\mbox{\raisebox{-.6ex}{$\stackrel{<}{\sim}$}}}
{\newcommand{\ga}{\mbox{\raisebox{-.6ex}{$\stackrel{>}{\sim}$}}}

Vigorous research in string theory in non-trivial backgrounds
in the past few years has yielded a number of novel
approaches for the construction and classification of string vacua.
One such method is known as string duality, or twisting procedure,
or $~O(d,d+n)~$ boosting (in presence of $n$ independent gauge fields)
\cite{ODD}-\cite{MS}. It employs the fact that
many string solutions posses toroidal symmetries associated with
the appearance of commuting translational Killing vectors  spanning
the tangent-space basis. At the level of the effective field theory on
target space, the symmetry takes a very simple form. One can rewrite
the effective action in terms of the independent degrees of freedom,
after dimensional reduction ~{\it a ~l\`a}~ Kaluza-Klein, and extract
the $~O(d,d+n)~$ group as a part of the full invariance group of the
theory. Although the action and the equations of motion  (the  $\beta$-function
equations) are invariant under this group, the initial conditions are not.
Thus under an $~O(d,d+n)~$ transformation a solution to the equations
of motion transforms, in principle, to a different solution. (One has to
remember that $~O(d,d+n)~$ transformations contain diffeomorphisms
and gauge transformations which have to be factored out, since they
don't change solutions \cite{SEN1}. )

In this letter, I will show that there exists another transformation which
leaves the $\beta$-function equations form-invariant, although the action and
the solutions are changed. It is  realized by shifting the
dilaton field and scaling  the lapse function,  and in its appearance is
similar
to a string field redefinition, which does not change the physics of the
fundamental string \cite{FRD}.  For this reason, and for the sake of brevity,
I will tentatively refer to it as a "field redefinition",
dropping the term "string" to indicate that in general they
are different. Therefore, in what follows, by field redefinition
it is meant a transformation in the above sense, while the label string
field redefinition refers to the conventional ambiguity of string theory.
This field redefinition is not completely arbitrary, for one must require that
the
transformation preserve the form of the lowest order $\beta$-function equations
while changing solutions. Viewed as a consistency condition, this imposes a
constraint on the field redefinition parameters. The constraint admits
non-trivial
solutions for all string configurations from the class considered here, namely
with all but one toroidal isometries. The field redefinitions which solve the
constraint are sufficiently general to produce new solutions. It is
particularly
interesting to note that among them there are field redefinitions which map
solutions with  string cosmological constant to those without. The role of such
field redefinitions is illustrated with several simple examples involving two-
and
three-dimensional black holes and black strings.

The effective action of string theory describing dynamics of the
massless bosonic background fields to the lowest
order in the inverse string tension $\alpha'$ expansion is, in the
world-sheet frame
\begin{equation}\label{1}
S~=~\int d^{d+1}x\sqrt{g}  e^{- \Phi}
\big(R +\partial_{\mu}\Phi \partial^{\mu} \Phi
-{1 \over 12} H_{\mu\nu\lambda}H^{\mu\nu\lambda}
-{\alpha' \over 4} F^{N}{}_{\mu\nu}F^{N}{}^{\mu\nu}
+ 2 \Lambda \big)
\end{equation}

\noindent The action above is written in Planck units $\kappa^2 = 1$.
Here $~F^{N}{}_{\mu\nu}= \partial_{\mu}A^{N}{}_{\nu} -
\partial_{\nu}A^{N}{}_{\mu}~$
are field strengths of $n$ Abelian gauge fields $A^{j}{}_{\mu}$,
$~H_{\mu\nu\lambda}= \partial_{\lambda}B_{\mu\nu} + ~cyclic~permutations~
- ({\alpha'/2}) \Omega_M{}_{\mu\nu\lambda}~$
is the field strength associated with the Kalb-Ramond
field $~B_{\mu\nu}~$ and $~\Phi~$
is the dilaton field. The $n$ Abelian gauge fields should be thought of as
the components of a non-Abelian gauge field ${\bf A}$ residing in the
Cartan subalgebra of the gauge group, while the
rest have been set equal to zero.
In this letter I will set $\alpha' = 1$.
The Maxwell Chern-Simons form
$\Omega_M{}_{\mu\nu\lambda} =
\sum_N A^{N}{}_{\mu}F^{N}{}_{\nu\lambda}
+ ~cyclic~permutations~ $ appears
in the definition of the axion field strength due to the Green-Schwarz anomaly
cancellation mechanism, and can be understood as a model-independent
residue after dimensional reduction from ten-dimensional superstring theory.
In fact, this term is a necessary ingredient of the theory if one wants to
ensure the
$O(d,d+n)$ invariance, as shown by Maharana and Schwarz \cite{MS}.

Under the assumption that all but one coordinates are toroidal, one can
dimensionally reduce the action (\ref{1}) to its effective
form describing dynamics of those stationary points with
$d$ commuting isometries. The reduction is a generalization of the
Kaluza-Klein dimensional reduction in presence of matter fields,
and is straightforward, if tedious \cite{ODD}-\cite{MS}. Below
I will briefly review some of the main ingredients, to the
extent necessary for the purpose of this letter.
I will work with the assumption that there are no
cross terms of the form $drdx^k$ in the metric and the
axion. In fact, it is easy to show that
when all but one coordinates are toroidal, this {\it is} the
most general ans\" atz, by employing gauge transformations
and diffeomorphisms. The cross terms $drdx^k$ in the metric corresponding to
the
"shift" functions can be removed by coordinate transformations
$x^k \rightarrow x^k + F^k(r)$. The remaining cross terms in the axion
field and the $r$ component of the gauge fields can be removed
by gauge transformations. To see that this is consistent with the equations
of motion for these modes, one just needs to recall that  they
are homogeneous for gauge degrees
and diffeomorphism and gauge invariant, and hence admit trivial solutions.
Therefore,  the backgrounds studied here will be of the form
\begin{eqnarray}\label{2}
ds^{2} &=& \Gamma(r) ~dr^{2}+G_{jk}(r)~dx^j dx^k \nonumber \\
B &=& \frac{1}{2}~ B_{jk}(r) dx^j \wedge dx^k \\
A^{N} &=& A^{N}{}_{j}(r) dx^j  \nonumber \\
\Phi &=& F(r)  \nonumber
\end{eqnarray}

\noindent where the $d \times d$ matrix $G_{jk}(r)$ is either of signature $d$
(in which case $\Gamma > 0$ and $r$ is the time coordinate) or of signature
$d-2$
(when $\Gamma < 0$ and $r$ is a space-like coordinate).
The lapse  function $\Gamma$ is kept arbitrary as its variation
in (\ref{1}) yields the constraint equation.

With this ans\" atz, the action (\ref{1}) can be rewritten in the
manifestly $O(d,d+n)$ invariant form:
\begin{equation}\label{3}
S_{eff}~=~\int d r \sqrt{\Gamma}  e^{-\phi}
\Big({1 \over \Gamma}\phi'^2
+ {1 \over 8\Gamma}Tr\bigl({\cal M}'{\cal L}\bigr)^2 + 2 \Lambda\Big)
\end{equation}

\noindent  where the prime denotes the derivative with respect to $r$.
One has to distinguish between the true dilaton $\Phi$
and the effective dilaton after dimensional reduction
$\phi = \Phi - (1/2) \ln |\det G |$. The two matrices
$~{\cal M}~$ and $~{\cal L}~$  which appear in the action (\ref{3}) are defined
by
\begin{eqnarray}\label{4}
&& {\cal M}~=~\pmatrix{
g^{-1}&-g^{-1}C&-g^{-1}A \cr
-C^{T}g^{-1}&g  + a + C^{T}g^{-1}C&A + C^{T}g^{-1}A \cr
-g^{-1}A&A + C^{T}g^{-1}A&{\bf 1} +  A^{T}g^{-1}A \cr}    \nonumber\\
&&~~~~~~~~  \\
&&~~~~~~~~~~~~~~~~~~~~~ {\cal L}~=~\pmatrix{~0~&{\bf 1}&~0~ \cr
{\bf 1}&~0~&~0~ \cr
{}~0~&~0~&{\bf 1} \cr} \nonumber
\end{eqnarray}

\noindent Here $~g~$ and $~b~$ are $~d \times d~$ matrices
defined by the dynamical degrees of freedom of the metric and the
axion:$~g~=~\bigl(G_{jk}\bigr)$ and $b~=~\bigl(B_{jk}\bigr)$. The matrix $A$ is
a $~d \times n~$ matrix built out of the gauge fields:
$A_{kN}~=~A^{N}{}_{k}$. The matrix $a$ is defined by
$a =  AA^{T}$, and $C$ by $C = (1/2)a + b$. The first two blocks $~{\bf 1}~$
in ${\cal L}$ are $~d \times d~$ matrices, and the last is $~n \times n~$. Note
that
${\cal M}^{T} = {\cal M}$ and
${\cal M}^{-1} = {\cal L}{\cal M}{\cal L}$. Therefore, ${\cal M}$ is a
symmetric element of $O(d,d+n)$. An invariance of the
theory then is obviously a chiral $O(d,d+n)$ rotation
${\cal M} \rightarrow  \Omega {\cal M}  \Omega^{T}$,
which preserves the symmetry property of ${\cal M}$.
The complete set of $\beta$-function equations can be obtained
from (\ref{3}) by the standard variational procedure \cite{VEN1}:
\begin{eqnarray}\label{5}
 \frac{\phi'^2}{\Gamma} +
\frac{1}{8\Gamma} Tr\bigl({\cal M}'{\cal L}\bigr)^2 &=& 2\Lambda\nonumber\\
 \bigl(\frac{e^{-\phi}}{\sqrt{\Gamma}}\phi'\bigr)' +  2\Lambda
\sqrt{\Gamma}e^{-\phi} &=& 0 \\
 \bigl(\frac{e^{-\phi}}{\sqrt{\Gamma}} {\cal M}'{\cal M}^{-1}\bigr)'  &=& 0
\nonumber
\end{eqnarray}

\noindent The ${\cal M}$ equation is readily integrable. Obviously \cite{VEN1},
\begin{equation}\label{6}
{\cal M}' = \sqrt{\Gamma} e^{\phi} {\cal J} {\cal M}
\end{equation}

\noindent We recognize ${\cal J}$ to be a restriction on ${\cal M}$ of the
Maurer-Cartan form for the group $O(d,d+n)$. As a consequence,  ${\cal J}$ is a
constant element of the Lie algebra
of $O(d,d+n)$, and satisfies two relations which in terms of
${\cal K} = {\cal J} {\cal L}$, can be written
as ${\cal K}^{T} = - {\cal K}$ and
${\cal K}{\cal L}{\cal M} = - {\cal M}{\cal L}{\cal K}$. Now one can compute
the
trace of $({\cal M}'{\cal L})^2$:
$Tr({\cal M}'{\cal L})^2 = - \Gamma \exp(2\phi)Tr{\cal J}^2 = -8\lambda \Gamma
\exp(2\phi)~$.
I have
introduced $\lambda = (1/8)Tr{\cal J}^2 = {\rm const.}$ as a short-hand
notation for this trace. The remaining equations
of (\ref{5}) can be rewritten as
\begin{eqnarray}\label{7}
 \frac{\phi'^2}{\Gamma}  &=& 2\Lambda  + \lambda e^{2\phi}\nonumber\\
 \bigl(\frac{e^{-\phi}}{\sqrt{\Gamma}}\phi'\bigr)' &=& -   2\Lambda
\sqrt{\Gamma}e^{-\phi}
\end{eqnarray}

\noindent They and the first integral (\ref{6}) comprise the full set of
equations
equivalent to the $\beta$-function system (\ref{5}). Actually,  the first of
Eq. (\ref{7}) is the first integral of the second, so in principle the second
equation can be ignored. However, it will be kept here as it will turn out to
be useful later.

Before solving these equations, it is advisable to inspect
if they admit any symmetries which
could be of interest. Indeed, the equations (\ref{6})-(\ref{7}) are of fairly
general nature, and
knowing any symmetries in addition to the $O(d,d+n)$ invariance could provide
us with
further insight into the structure of the solution space of the theory. The
requirement of
symmetry can be somewhat relaxed. One can ask if there are invariances of the
$\beta$-function equations (i.e., their first integrals  (\ref{6})-(\ref{7}))
which leave the form of the $\beta$-functions the same but change the action
(\ref{3}) and the solutions. In this case, all
one needs to require is that the realization of this invariance must map
solutions on solutions. That there is an additional invariance of the system
(\ref{6})-(\ref{7}) of this type can be seen as follows. Let the dilaton and
the lapse
simultaneously undergo a shift and scaling according to
\begin{equation}\label{8}
\phi \rightarrow \tilde \phi = \phi + \chi  ~~~~~~~~~
\Gamma \rightarrow \tilde \Gamma = \Gamma \exp(-2\chi)
\end{equation}

\noindent while the "matter" ${\cal M}$ is left
unchanged; obviously, the equation (\ref{6}) is invariant
under this transformation. Note that this implies that the metric, axion and
gauge fields in the
subspace of the original $d+1$-dimensional manifold spanned by the Killing
vectors also remain unchanged. To assure that $\tilde \chi, \tilde \Gamma$ are
solutions one must
impose the form-invariance of  (\ref{7}) under the transformation  (\ref{8}).
This will be true if the field redefinition parameter is constrained to satisfy
\begin{eqnarray}\label{9}
\chi'^2 + 2 \phi'  \chi' - 2\Gamma
\Bigl( \tilde \Lambda  e^{-2\chi} - \Lambda \Bigr) &=& 0\nonumber  \\
 \bigl(\frac{e^{-\phi}}{\sqrt{\Gamma}} \chi' \bigr)' +
 2 \sqrt{\Gamma}e^{-\phi} \Bigl(\tilde \Lambda  e^{-2\chi} - \Lambda\Bigr) &=&
0
\end{eqnarray}

\noindent Here $\tilde \Lambda$ appears since  (\ref{6})-(\ref{7}) was required
to be form-invariant, which is more relaxed. This shows that $\tilde \chi,
\tilde \Gamma$ solve
 (\ref{6})-(\ref{7}) with a different value of the cosmological constant. As a
consequence,
if there exist solutions of (\ref{9}) with $\tilde \Lambda \ne \Lambda$, the
transformations (\ref{8})
generated by them will represent maps between solutions with different
cosmological
constant.  Moreover, using (\ref{9}) it is easy to see that the action changes
by
just a boundary term:
$~\delta S = \tilde S_{eff} - S_{eff} = 2 \int dr [(\exp(-\phi) /
\sqrt{\Gamma}) \chi']'$. Therefore,
the field redefinition (\ref{8}) represents a symmetry of the action in the
generalized sense,
since the variation is reduced to the boundary.

At this point one needs to solve the constraint (\ref{9}), and see if it admits
solutions other then the trivial, $\chi = 0$. Recalling that the lapse $\Gamma$
is a gauge parameter,  one can fix it to be $\Gamma = \exp{2\phi}$. Then
Eq. (\ref{9}) reduce to
\begin{eqnarray}\label{10}
\chi'^2 + \frac{\Gamma'}{\Gamma} \chi' &=& 2\Gamma
\Bigl( \tilde \Lambda  e^{-2\chi} - \Lambda \Bigr) \nonumber\\
 \chi'' + \chi'^2 &=&0
\end{eqnarray}

\noindent  The simplicity of the second equation has been hinted at earlier by
retaining
the second order differential equation in (\ref{7}). The general non-trivial
solution
is $~\chi = \chi_0 + \ln(r + \varpi)~$ with the two integration constants
$\chi_0$ and $\varpi$
constrained by the first of Eq. (\ref{10}). In order to see if there are
any allowed values of the parameters, one
needs to find $\Gamma$. In the gauge chosen for it, the system (\ref{7}) also
simplifies
greatly, and the most general solution for $\Gamma$ is
\begin{equation}\label{11}
\Gamma  = \Bigl(2\Lambda r^2 + \omega r + \frac{\omega^2}{8\Lambda}
- \frac{\lambda}{2\Lambda} \Bigr)^{-1}
\end{equation}

\noindent which is correct even when $\Lambda \rightarrow 0$ (as can be seen
from (\ref{7}),
$\omega^2 \rightarrow 4\lambda + \Lambda\eta_0$,
so there exists a non-singular limit). Substituting this
in the first of Eq. (\ref{10}) gives the algebraic form of the constraint
(\ref{9}). It is
\begin{equation}\label{12}
4 \Lambda \varpi  = \omega \pm \sqrt{4\lambda
+ 16\Lambda \tilde \Lambda e^{-2 \chi_0} }
\end{equation}

\noindent After determining those parameters which solve (\ref{12}), the new
solution
can be obtained from (\ref{2}) and (\ref{8}), keeping in mind that the metric,
axion and gauge fields in the subspace spanned by the Killing vectors do not
change (and therefore
the physical dilaton $\Phi$ changes in the same way as the effective dilaton
$\phi$
because $|\det G |$ is invariant) :

\vfill
\newpage

\begin{eqnarray}\label{13}
ds^{2} &=&e^{-2\chi(r)} \Gamma(r) ~dr^{2}+G_{jk}(r)~dx^j dx^k \nonumber \\
B &=& \frac{1}{2}~ B_{jk}(r) dx^j \wedge dx^k \\
A^{N} &=& A^{N}{}_{j}(r) dx^j  \nonumber \\
\Phi &=& F(r) + \chi(r) \nonumber
\end{eqnarray}

 It is obvious that in general (\ref{13}) will represent
a solution different from (\ref{2}) if $\chi$ exists. There
still remains the question if this solution will also correspond to a {\it
different} physical situation.
That there might be such an ambiguity can be noted from the similarity between
the field redefinition (\ref{8}) and a special type of the string field
redefinition, defined by
(for simplicity,  in the absence of gauge fields):
\begin{eqnarray}\label{14}
G_{\mu\nu} \rightarrow \tilde G_{\mu\nu} &=& G_{\mu\nu}
+ S\bigl(R, (\partial \Phi)^2\bigr) \partial_{\mu} \Phi  \partial_{\nu}
\Phi\nonumber \\
\Phi \rightarrow \tilde \Phi &=&\Phi + T\bigl(R, (\partial \Phi)^2\bigr)
\end{eqnarray}

\noindent with the axion unchanged. In the case studied here, since all the
fields depend
only on one variable $r$, the equations (\ref{8}) would be of the form
(\ref{14}), if
$\chi = T = -(1/2) \ln[1 + S  (\partial \Phi)^2]$. To determine whether the
field redefinition (\ref{8}) is precisely the same as (\ref{14}), one needs to
restore the $\alpha'$ dependence in the action and the solutions and then
evaluate $\chi$. However, since $\chi$ relates solutions with different
cosmological constants, it is hard to see how in general it would be possible
to write it as an analytic function of  $R$ and $(\partial \Phi)^2$ only. I
hope to return to this issue
elsewhere.

In order to illustrate the arguments above, I will present several examples
involving two- and three-dimensional black holes, and investigate the effect of
the field redefinition (\ref{8}) on these solutions. The two-dimensional
solutions are the simplest possible. There are several
apparently different two-dimensional stringy solutions,  related to the Witten
two-dimensional
black hole \cite{EW} either by an $O(1,2)$ duality or
by extending the Witten solution to include all the higher order $\alpha'$
corrections. Actually, as noted  recently by Tseytlin \cite{TsFD},
the extension is related to the original solution by a string field
redefinition, and therefore to
distinguish between them we need to investigate dynamics of other fields
in these backgrounds (e.g., tachyon \cite{TsFD}).
On the other hand, it is well known that the $2d$ black hole extended  to all
orders in $\alpha'$
is non-singular \cite{PTY}. This indicates that
the singularity of the Witten solution is not as dangerous as are the
singularities in conventional General Relativity, because it disappears from
one version of the two solutions. It is
introduced in the metric
by the particular form of the string field redefinition, or in other words, it
is related to the specific subtraction scheme adopted in string field theory.
In its semiclassical singular form, the Witten $2d$ black hole is given by
\begin{eqnarray}\label{15}
ds^{2} &=&\frac{1}{2\Lambda} ~\frac{dr^{2}}{r(r-\mu)}
- \bigl(1 - \frac{\mu}{r} \bigr) dt^2 \nonumber \\
\Phi &=& - \ln \sqrt{2 \Lambda} r
\end{eqnarray}

\noindent Since $\phi = \Phi - (1/2)| \det G|$, and so $\exp(2\phi) = 2\Lambda
r(r-\mu)$, this is
precisely the gauge in which the detailed form of the field redefinition
(\ref{8}) is worked out.
For this solution, $\omega = -2 \Lambda \mu$ and $\lambda = \Lambda^2 \mu^2$.
The constraint (\ref{12}) then reads
\begin{equation}\label{16}
\varpi  =  - \frac{1}{2} \mu \pm \frac{1}{2}\sqrt{\mu^2
+ 4 \frac{ \tilde \Lambda}{\Lambda} e^{-2 \chi_0} }
\end{equation}

\noindent In the case when $\tilde \Lambda \ne 0$, the values of cosmological
constants
and $\chi_0$ combine into a single parameter, and (\ref{16}) can be inverted to
give
$\tilde \Lambda =  \Lambda \varpi (\varpi + \mu) \exp(2\chi_0)$.  I will skip
the details and just state that the redefined solution actually turns out to be
identical to (\ref{15}), after a coordinate transformation, as can be verified
by some straightforward algebra. The only interesting cases then are when
$\tilde \Lambda = 0$. Then either $\varpi = - \mu$ or $\varpi = 0$. The field
redefinition by
$\chi = \chi_0 + \ln(r - \mu)$ gives, after some manipulation,  the dual of the
Rindler solution in two dimensions
\begin{eqnarray}\label{17}
ds^{2} &=& dz^2
- \frac{ dt^2}{z^2} \nonumber \\
\Phi  &=&  \Phi_0 - 2 \ln z
\end{eqnarray}

\noindent which is related by Wick rotation to the singular $2d$ cosmology
discovered by Veneziano \cite{ODD}.
The other field redefinition,  $\chi = \chi_0 + \ln r $, gives, as one should
guess, the dual of (\ref{17}), i.e. the $2d$ Rindler solution:
\begin{eqnarray}\label{18}
ds^{2} &=& dz^2
- z^2 dt^2 \nonumber \\
\Phi &=& \Phi_0
\end{eqnarray}

One may wonder why these field redefinitions have not reproduced the result of
Tseytlin,
i.e. gave the exact form of (\ref{15}) to all orders in $\alpha'$.  This is
because the constraint
(\ref{9}) was enforced on the field redefinition, so it does not change the
form of
(\ref{6})-(\ref{7}), whereas the string field redefinition used in \cite{TsFD}
changes the equations
of motion by introducing corrections of higher order in $\alpha'$. Thus the two
solutions do not solve the same equations although they represent the same
physical situation.

In three dimensions, it is most convenient to investigate the action of
(\ref{8}) on the black string
solution of Horne and Horowitz \cite{HorHo}. It is not difficult to see that
all field redefinitions
with $\varpi$ different from both $-\mu$ and $-q^2/\mu$ and with $\tilde
\Lambda \ne 0$ do
not change the form of the black string, much the same way as in the
two-dimensional
black hole example above. For brevity, I will consider in detail only the two
special  field redefinitions which change the solution.
The black string solution is, in the gauge adopted here, given by
\begin{eqnarray}\label{19}
ds^{2} &=&\frac{1}{2\Lambda} ~\frac{dr^{2}}{(r - (q^2/\mu) )(r-\mu)}
+  \bigl(1 - \frac{q^2}{\mu r} \bigr) dx^2 - \bigl(1 - \frac{\mu}{r} \bigr)
dt^2 \nonumber \\
B &=& \frac{q}{r}~dx \wedge dt \\
\Phi &=& - \ln \sqrt{2 \Lambda} r \nonumber
\end{eqnarray}

\noindent Here,  $\omega = -2 \Lambda \mu (1 + (q/\mu)^2)$ and
$\lambda = \Lambda^2 \mu^2 (1 - (q/\mu)^2)^2$.  The constraint (\ref{12})
becomes
\begin{equation}\label{20}
\varpi  = -\frac{1}{2} \mu (1 + \frac{q^2}{\mu^2} ) \pm
\frac{1}{2} \sqrt{\mu^2 (1 - \frac{q^2}{\mu^2} )^2
+ 4\frac{ \tilde \Lambda}{\Lambda} e^{-2 \chi_0} }
\end{equation}

\noindent Let $\varpi =0$  and
$\tilde \Lambda = \Lambda$ (then, $\exp(-2\chi_0) = q^2$).
The transformed solution is, in terms of  the new coordinates
$z = (\mu - r)/\mu r$, $t = \vartheta /\sqrt{\mu}$ and $x = \sqrt{\mu} \tau
/q$,
with $m = (\mu^2 - q^2)/q^2 \mu$ and after a gauge shift
of the axion $B \rightarrow B - d(\theta d \vartheta) /\mu$,
found to be exactly the stringy version \cite{HWK} of the static
three-dimensional black hole \cite{BTZ}:
\begin{eqnarray}\label{21}
ds^{2} &=&\frac{1}{2\Lambda} ~\frac{dz^{2}}{z(z - m)}
+  z d\vartheta^2 - (z - m)d\tau^2 \nonumber \\
B &=& z~d\tau \wedge d\vartheta \\
\Phi &=&  \Phi_0  \nonumber
\end{eqnarray}

\noindent  It has been demonstrated before that the three-dimensional black
string was
related to the three-dimensional spinning black hole by a duality
transformation. With the
field redefinition approach one can relate the black string to the {\it static}
three-dimensional
black hole. Note that this is the only way to field-redefine the solution so
that the dilaton is constant when the axion is present. When $q = 0$, the
solution reduces to the Witten black hole extended with a flat coordinate, and
then it is possible to map it to a flat solution in three  dimensions, as
illustrated by the argument leading to (\ref{18}).

In the last example,  let the black string (\ref{19}) be extremal: $q = \mu$.
Then,
if $\tilde \Lambda = 0$, the constraint (\ref{20}) admits the unique solution
$\varpi = -\mu$.
After some manipulation, the new solution can be put in the form of a singular
configuration of the type studied in \cite{COH}:
\begin{eqnarray}\label{22}
ds^{2} &=&dz^{2}  +  \frac {d\vartheta^2 - d\tau^2}{z} \nonumber \\
B &=& \frac{1}{z}  d\tau  \wedge d\vartheta \\
\Phi &=&  \Phi_0 - \ln z \nonumber
\end{eqnarray}

To summarize, it is clear that the field redefinition method presented in this
letter
can be useful in the study of string solutions. It is a concise and practical
tool for the
construction of semiclassical string vacua. While not a substitute for the
powerful
method of string duality,  in the simplest situations exhibited
here, their usefulness is perhaps compatible.
In fact, its value and versatility are enhanced when  used in conjunction with
duality.
The method provides a supplement to duality because it
can be used to relate solutions with different values of the string
cosmological constant.
This offers further means to distinguish between the roles
played by what we perceive as the
cosmological constant in General Relativity and in string theory.
It has been proposed recently in the first reference of \cite{HWK} and in
\cite{CC} that the status of the string cosmological constant
must be fundamentally different
from  the status of the cosmological constant in General Relativity.
The arguments were based on the
observation that string duality in some cases relates asymptotically flat
solutions
to those which are not asymptotically flat, although both solve the
$\beta$-function equations with the string cosmological term (e.g. the  $3d$
black string - $3d$ black hole duality).
The field redefinitions described here support this observation, as
they provide direct means to connect solutions with different string
cosmological term.
Furthermore, the field redefinitions  could also be useful when combined with
the
singular limits approach, another unorthodox method for generating
string solutions \cite{KUM}.

It would be interesting to see how this procedure generalizes on
backgrounds which have less isometries then what was assumed here, i.e. with
two or more
non-toroidal coordinates.  In fact, it is not very hard to conclude that the
procedure will
admit rather straightforward generalization to the problems with three or more
non-toroidal
coordinates, on the basis of the form of the dimensionally reduced action in
those
cases.  For the problems with two non-toroidal coordinates (which encompass
such physically interesting situations as four-dimensional stationary
spherically symmetric
backgrounds) the issue is more subtle and needs to be addressed with more care.
If it turns
out to be possible to adapt the method described here to such cases, then we
may be able
to use it with the string version of the Geroch group-extended duality  and
thus enhance
both approaches \cite{BAK}.
\vskip1cm

{\bf Acknowledgements}
\vskip0.5cm
I would like to thank  B. Campbell, G. Hayward and K. Olive
for interesting conversations. K. Sfetsos was kind to point out
an oversight, for which I thank him. I would also like to thank
the Aspen Center for Physics, where a part of the research reported here has
been
done,  for kind hospitality. This work has been supported in part by
the Natural Science and Engineering Research Council of Canada.

\vfill
\eject

\end{document}